\documentclass[a4paper,11pt]{article}
\usepackage[utf8]{inputenc}
\usepackage{multicol}

\usepackage{graphicx}
\usepackage[dvipsnames]{xcolor}
\usepackage{graphicx,times}             
\usepackage{natbib}
\usepackage{amssymb,amsmath}

\DeclareUnicodeCharacter{2212}{\textendash}
\bibpunct{(}{)}{;}{a}{}{,}

\usepackage[pagebackref=true]{hyperref}

        \title{Maximal mass of the neutron star with a deconfined quark core}

        \author{Muhammed Shafeeque$^1$\footnote{m.shafeeque@iitg.ac.in}, Arun Mathew$^2$\footnote{arun@cp.dias.ie} and Malay K. Nandy$^1$\footnote{mknandy@iitg.ac.in}\\
        $^1$\small Department of Physics, Indian Institute of Technology Guwahati, Guwahati 781039, India.\\
        $^2$\small Dublin Institute for Advanced Studies, Astronomy \& Astrophysics Section, Dublin D15 XR2R,
             Ireland.}

        \date{\vspace{-5ex}}

\begin{document}

\maketitle

\begin{abstract}
The nature of equation of state for the matter in the neutron star plays an important role in determining its maximal mass. In addition, it must comply with the condition of causality. Noting that the central density of a maximally massive neutron star is well above the nuclear saturation density, a deconfined quark core in the central region is motivated in this paper. We analyze this scenario by employing the MIT bag model to represent the core region and one of the unified equations of state for the region outside the core. Such combination is found to solve the problem of causality violation. In each case of the combined equations of state, the radial profile of $\rho r^2$ displays a peak and dominant contribution to the total mass of the star comes from the region around the peak value of $\rho r^2$, whereas the contribution is small from the regions near the center and the surface. This peak occurs in the region of hadronic matter for the combinations considered in this paper. Importantly, we find that the position of the peak in $\rho r^2$ is well-correlated with the maximal mass--- the highest value of $1.98\ M_\odot$ obtains for the case with the peak occurring farthest from the center. This gravitational threshold being obtained for a non-rotating neutron star, we expect the threshold to lie well above 2 $ M_\odot$ for a rapidly rotating neutron star, that may explain the existance of massive pulsars from recent astronomical observations.
\end{abstract}

\section{Introduction}\label{sec_intro}

The earliest proposal for the possible existence of a dense star, such as a neutron star, was made by \cite{landau1932}. Later, \cite{tolman} and \cite{opp_volk}, employing the  celebrated Tolman-Oppenheimer-Volkoff (TOV) equations, obtained the stellar structure of  neutron stars in general relativity (GR). Neutron stars became a topic of interest after the discovery of X-ray sources by \cite{Giacconi1962} and the discovery of pulsar in 1967 by \cite{HEWISH1968}. Pulsars were then identified as rotating neutron stars by \cite{GOLD1968, GOLD1969}. Binary neutron star was first discovered in 1975 by \cite{Hulse1975}, and the millisecond pulsar (MSP) by \cite{Backer1982}. In all subsequent observations until before 2010, the pulsars were found to have masses well below $2\ M_\odot$.

However, the scenario underwent a remarkable change since 2010 when pulsars of stellar mass $\sim 2$ $M_\odot$ were observed to exist in the Universe. For instance, \cite{Demorest2010} discovered MSP J1614-2230 in 2010 having a pulsar mass of $1.928\pm0.017$ $ M_{\odot}$ (as re-estimated by \cite{Fonseca_2016}). Similarly, \cite{Antoniadis2013} discovered MSP J0348+0432 in 2013 with a pulsar mass of $2.01\pm0.04$ $ M_{\odot}$. In recent years, massive pulsars have been detected such as PSR J2215+5135 by  \cite{Linares2018} with a pulsar mass of $2.27\pm_{0.17}^{0.15}$ $ M_{\odot}$, MSP J0740+6620 by \cite{Cromartie2020} with a pulsar mass of $2.14\pm_{0.09}^{0.10}$ $ M_{\odot}$, and PSR J0952-0607 by \cite{PSRJ095206071} with a pulsar mass of $2.35\pm0.17$ M$_\odot$ \cite{PSRJ095206072}.

Recent gravitational wave detections also indicated the existence of massive neutron stars. The gravitational wave event GW170817  (\cite{GW170817}), the first observed binary neutron star merger by the LIGO, reported component masses in the range $1.0-1.89$ $M_\odot$ with a total mass of $2.73^{+0.04}_{-0.01}$ $M_\odot$  (\cite{GW170817a}). The next observed binary neutron star merger event GW190425  (\cite{GW190425}) reported component masses between $1.15-2.52$ $M_\odot$ with a total mass of $3.4^{+0.3}_{-0.1}$ $M_\odot$. Moreover, the GW190814 event was predicted to be the merger of a massive compact object of $2.50-2.67$ $M_\odot$ with a black hole of $22.2-24.3$ $M_\odot$  (\cite{GW190814}). The compact object, although too massive to be a neutron star and too light to be a black hole, might as well be a candidate for a massive neutron star.

The observational evidences for massive neutron stars naturally require a theoretical explanation including their stellar structure and composition of matter interior to such massive objects. However, such evidences bring about constraints on the equation of state, so that a stellar mass of $\sim 2$ $M_\odot$ could be obtained (\cite{massradii}). Consequently, efforts have been made to obtain an equation of state with a high degree of stiffness. In fact, high stiffness was incorporated for the high density regions of the star by removing the hyperons from the nuclear matter  (\cite{nohyp1, nohyp2, fermiliqeos}). However, it was argued that hyperons would emerge in the high density nuclear matter leading to softening of the equation of state  (\cite{hyperon1,hyperon2}).

There have been attempts to formulate the equation of state of neutron matter considering multi-body interactions employing various phenomenologies such as Skyrme-type forces ( \cite{SK0, SK1,  FPS1, FPS0, SLy_Chabanat_Haensel_1, SLy_Chabanat_Haensel2}), relativistic mean field ( \cite{RMF1,RMF2,RMF3, RMF4}), chiral effective field theory  (\cite{EFT}), the Urbana model  (\cite{APR}), relativistic Fermi liquid  (\cite{fermiliqeos}), etc. Some of these equations of state are stiff enough to yield the maximum stable mass $M_{\rm max}\sim2$ $M_\odot$ or even higher. However, some of these equations of state violate the condition of causality in the high density region of the star.

Moreover, there exist formulations, usually referred to as unified equations of state, such as SLy4, FPS and BSk $19-21$ (\cite{SLy_Haensel1_Potekhin,bsk_Potekhin}). These unified equations of state are based on Skyrme type nucleon-nucleon interactions. SLy4 was constructed from SLy (Skyrme Lyone) type interactions  (\cite{SLy_Chabanat_Haensel_1, SLy_Chabanat_Haensel2}) and FPS was from generalized Skyrme model  (\cite{FPS0}) by fitting to the estimates of Ref.\  (\cite{FPS1}). BSk$19-21$ were based on the corresponding energy density functionals (EDF), developed from generalized Skyrme force using the Hartee-Fock-Bogoliubov method by the Brussels-Montreal group  (\cite{bsk_Goriely, bsk_Pearson, bsk_Pearson_Goriely_chamel}). BSk19 EDF was fitted with constraints from the equation of state in Ref.\  (\cite{FPS1}). BSk20 EDF was fitted with constraints from APR equation of state  (\cite{APR}), whereas BSk21 EDF was fitted with constraints from LS2 equation of state  (\cite{LS2}). It may be noted that these hadronnic equations of state are fitted with constraints coming from standard equations of state which have been very successful in modelling neutron stars.

These hadronic equations of state (SLy, FPS, and BSk19--21) have certain advantageous features. Each of these equations of state is formulated using the same many-body interaction so that transitions between different hadronic regions of the star are continuous. They also remain causal up to reasonably high densities.

 Among the five unified equations of state, SLy, BSk20 and BSk21 satisfy the 2~M$_\odot$ constraint coming from observations of massive pulsars (\cite{Demorest2010,Fonseca_2016,Antoniadis2013,Linares2018,Cromartie2020}), as they are capable of yielding maximal stable mass $M_{\rm max}$ $\sim 2$ $M_\odot$ or greater, BSk21 giving the highest  (\cite{SLy_Haensel1_Potekhin,bsk_Potekhin}).

 As we shall see in Section  \ref{sec_ic} below, when some of these hadronic equations of state (BSk19 and BSk20) are employed throughout the star, they violate the condition of causality ($c_s\leq c$) just before the gravitational instability sets in, so that, at ultra-high densities in the core region, the speed of sound $c_s$ exceeds the speed of light $c$. The remaining hadronic equations of state (Sly, FPS, BSk21) remain causal within their regions of gravitational stability.

 It is also important to note that, when these hadronic equations of state (SLy, FPS, and BSk19--21) are employed throughout the star, in order to attain the maximal mass configurations ($M_{\rm max}\sim2\ M_\odot$), the central densities turn out to be $\rho_c\sim10\ \rho_{\rm nuc}$, where $\rho_{\rm nuc}=2.67\times 10^{14}$ g cm$^{-3}$ is the nuclear saturation density. Quark deconfinement is expected to occur at such ultra-high densities in the core region (\cite{hydrostatic_MIT_Naok, quark_stars_Ivanenko, SHURYAK}).

 The above facts indicate that hadronic equations of state are not adequate representations for the ultra-high density core region. On the other hand, they are most suitable to represent the outer regions of lower densities (where they are causal).

 In this paper, we thus motivate the scenario of deconfined quark matter in the core region and hadronic matter in the region outside the core. For a simplistic analysis of this scenario, we employ the MIT bag model   (\cite{MIT_jeff,MIT_Yu.A.Simonov}) to represent the deconfined quark matter in the core region and one of the unified equations of state (SLy4, FPS and BSk $19-21$   \cite{SLy_Haensel1_Potekhin,bsk_Potekhin}) for the hadronic matter in the outer region. With this construction, we find that the hadronic equations of state remain causal because of crossover to the MIT bag model equation of state, so that causality is maintained throughout the star.

 While there is no concrete result for the hadron-to-quark transition in the literature, \cite{pqcd} suggested that typical models predict the phase transition to occur between $2-10$ $\rho_{\rm nuc}$. On the other hand, \cite{Baym_2018} suggests that quark matter equation of state should be applied for densities higher than $\sim4-7$ $\rho_{\rm nuc}$.

 We find that these bounds are satisfied when we take the first crossover between the MIT bag model and the unified equations of state to determine the transition from the quark region to the hadronic region. Namely, the transition points occur at 2.92 $\rho_{\rm nuc}$, 4.10 $\rho_{\rm nuc}$, 4.72 $\rho_{\rm nuc}$, 6.33 $\rho_{\rm nuc}$ and 6.78 $\rho_{\rm nuc}$ for transition from the MIT bag model to BSk21, BSk20, SLy, BSk19 and FPS, respectively. Such piece-wise continuous crossover happens in a small region so that we do not expect a significant change in the {\em total} mass if this cross-over is made smooth by interpolation. With this scheme, we study the stellar structure and obtain the maximal stable mass $M_{\rm max}$ for central densities  $\rho_c\sim10\ \rho_{\rm nuc}$. Our analysis reveals interesting features in relation to the maximal mass within this framework.

 The remainder of the paper is organized as follows. Section  \ref{sec_field_eq} briefly outlines the field equations. Section  \ref{sec_eq_state} presents the analytical representations of the unified equations of state and the MIT bag model. In Section  \ref{sec_ic}, we give an outline of the methodology adopted in numerical integration of the field equations together with the boundary conditions. Section  \ref{sec_stellar_str} embodies the main motivation of this paper, with details of the stellar structure predicted by different combinations of the MIT bag model with the unified equations of state. Finally, the paper is concluded in Section  \ref{sec_concl} with a discussion of the results.

\section{Field Equations}\label{sec_field_eq}
In general relativity, the Einstein-Hilbert action is expressed as
\begin{equation}
 \mathcal{S}=\frac{1}{2c\kappa}\int d^{4}x\,\sqrt{-g}\,R+\int d^{4}x\,\sqrt{-g}\,\mathcal{L}_{m},\label{Einstein_hilbert_actio}
\end{equation}
 where $\kappa=\frac{8\pi G}{c^4}$, with $G$ the Newton's constant, $R$ the scalar curvature, and $\mathcal{L}_m$ is the matter Lagrangian. Varying the action with respect to the metric $g^{ab}$,  and applying Hamilton's principle, the Einstein field equations are obtained as
\begin{equation}
 G_{ab}=R_{ab}-\frac{1}{2}g_{ab}R=\kappa T_{ab},\label{Einstein equation}
\end{equation}
where $R_{ab}$ is the Ricci tensor, and $T_{ab}=cg_{ab}\mathcal{L}_{m}-2c\frac{\delta\mathcal{L}_{m}}{\delta g^{ab}}$ is the energy-momentum tensor.

For a perfect fluid, $T_{ab}=\frac{\left(\varepsilon+P\right)}{c^{2}}u_{a}u_{b}+g_{ab}P$, where $\varepsilon=\rho c^2$ is the energy density, $P$ is the pressure, and $u^a$ is the four-velocity with $u^au_a=-c^2$.

Using the spherically symmetric static metric, $ds^{2}=-e^{\nu(r)}\,c^{2}\,dt^{2}+e^{\lambda(r)}\,dr^{2}+r^{2}d\theta^{2}+r^{2}\sin^{2}\theta \,d\varphi^{2}$, leads to the field equations
\begin{equation}
 \begin{array}{cc}
  \displaystyle\frac{d\lambda}{dr}&=\displaystyle\frac{1-e^{\lambda}}{r}+\kappa re^{\lambda}\varepsilon(r),\\
  &\\
  \displaystyle\frac{d\nu}{dr}&=\displaystyle\frac{e^{\lambda}-1}{r}+\kappa re^{\lambda}P(r),
 \end{array}\label{field equation}
\end{equation}
where  $\lambda(r)$ and $\nu(r)$ are the undetermined metric potentials. The radial profile of mass can be obtained from
\begin{equation}\label{mass_profile_eqn}
  m(r)=\frac{c^2}{2G}r\{1-e^{-\lambda(r)}\}.
\end{equation}

From equations ( \ref{field equation}), ( \ref{mass_profile_eqn}), and the conservation law, $\nabla_{a}T^{ab}=0$, TOV equations are obtained as

\begin{equation}\label{TOV1}
 \displaystyle\frac{dP}{dr}=\displaystyle\frac{-\left[P(r)+\varepsilon(r)\right]}{r\left[c^2r-2Gm(r)\right]}\left[Gm(r)+\frac{4\pi G}{c^2} r^3P(r)\right],
\end{equation}

\begin{equation}\label{TOV2}
 \displaystyle\frac{dm(r)}{dr}=\displaystyle\frac{4\pi r^2\varepsilon(r)}{c^2},
\end{equation}



We integrate the above equations simultaneously for the interior as well as the exterior of the neutron star with appropriate boundary conditions (as discussed in Section  \ref{sec_ic}) employing different equations of state (as given in Section  \ref{sec_eq_state}).

\section{Equations of State}\label{sec_eq_state}
The nuclear matter in a neutron star is subjected to extreme conditions as it is under tremendous pressure with the central density $\rho_c\sim 10^{15}$ g cm$^{-3}$ for maximal mass configurations. This makes the nuclear matter highly degenerate and the equation of state is independent of temperature  (\cite{SLy_Haensel1_Potekhin}), except for a thin outer layer. The core is expected to consist of deconfined quark matter which we shall represent by the MIT-bag model  (\cite{MIT_jeff, MIT_Yu.A.Simonov}). The hadronic matter in the outer region will be represented by a unified equation of state such as, SLy, FPS  (\cite{SLy_Haensel1_Potekhin}), BSk19, BSk20 and BSk21  (\cite{bsk_Potekhin}.
)
 \begin{figure}
 \centering
 \includegraphics[width=.8\textwidth, height=7cm]{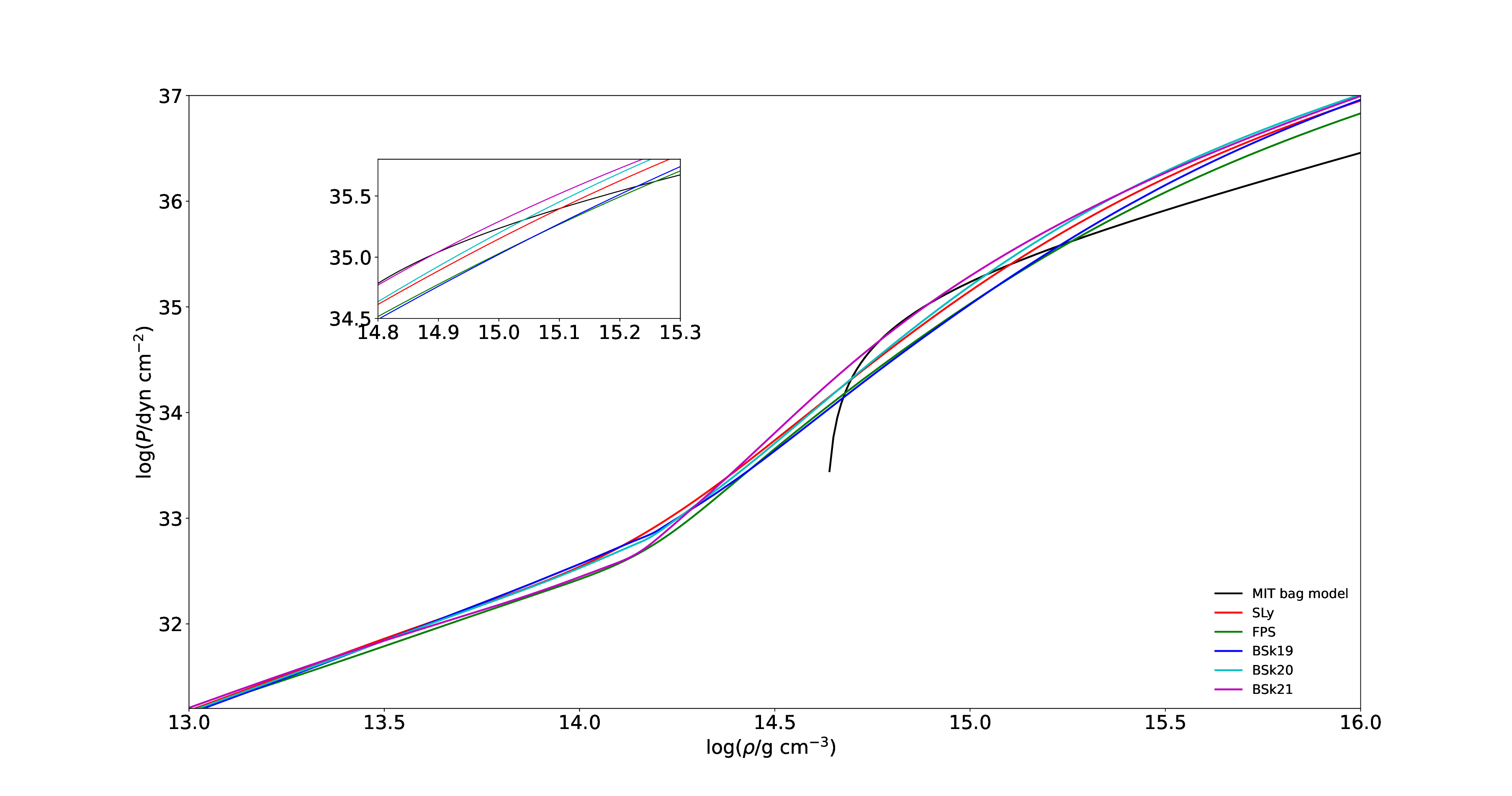}
 \caption{ \footnotesize Plots for different unified equations of state expressed by equations  \ref{SLy-FPS_eqn} and  \ref{bsk_eqns}, and the MIT bag model given by equation  \ref{mit_bag_model}. The inset shows the crossover points between the MIT bag model and the unified equations of state.}
 \label{fig.eos}
\end{figure}

In order to have continuous representations for these equations of state, we employ the parameterized forms of the pressure $P$ as a function of density $\rho$. The parameterized form in the SLy and FPS unified equations of state is expressed as  (\cite{SLy_Haensel1_Potekhin}),
\begin{equation}\label{SLy-FPS_eqn}
 \begin{split}
  &\zeta=\frac{a_{1}+a_{2}\xi+a_{3}\xi^{3}}{(a_{4}\xi+1)\left(e^{a_{5}(\xi-a_{6})}+1\right)}\\
  &\ \ +\frac{a_{7}+a_{8}\xi}{e^{a_{9}(a_{10}-\xi)}+1}+\frac{a_{11}+a_{12}\xi}{e^{a_{13}(a_{14}-\xi)}+1}\\
  &\ \ +\frac{a_{15}+a_{16}\xi}{e^{a_{17}(a_{18}-\xi)}+1},
 \end{split}
\end{equation}
where $\zeta=\log\,(P/P_{\rm ref})$ with $P_{\rm ref}=1.0$ dyne cm$^{-2}$, and $\xi=\log\,(\rho/\rho_{\rm ref})$ with $\rho_{\rm ref}=1.0$ g cm$^{-3}$. The eighteen parameters $a_i$ are given in Ref.  (\cite{SLy_Haensel1_Potekhin}).

Additionally, the parameterized form of the pressure $P$ as a function of density $\rho$ for BSk19, BSk20, and BSk21 unified equations of state is given by the expression  (\cite{bsk_Potekhin})
\begin{equation}\label{bsk_eqns}
 \begin{split}
  &\zeta=\frac{a_{1}+a_{2}\xi+a_{3}\xi^{3}}{\left(a_{4}\xi+1\right)\left(e^{a_{5}(\xi-a_{6})}+1\right)}+\frac{a_{7}+a_{8}\xi}{e^{a_{9}(a_{6}-\xi)}+1}\\
  &\ \ +\frac{a_{10}+a_{1}\xi}{e^{a_{12}(a_{13}-\xi)}+1}+\frac{a_{14}+a_{15}\xi}{e^{a_{16}(a_{17}-\xi)}+1}\\
  &\ \ +\frac{a_{18}}{\left[a_{19}(\xi-a_{20})\right]^{2}+1}+\frac{a_{21}}{\left[a_{22}(\xi-a_{23})\right]^{2}+1},
 \end{split}
\end{equation}
 with $\zeta=\log\,(P/P_{\rm ref})$ and $\xi=\log\,(\rho/\rho_{\rm ref})$, as before. The twenty-three parameters $a_i$ are given in Ref.  (\cite{bsk_Potekhin}).

 For the deconfined quark core, we employ the equation of state given by the MIT bag model  (\cite{MIT_jeff,MIT_Yu.A.Simonov}),
\begin{equation}\label{mit_bag_model}
 P=k(\rho c^2-4B),
\end{equation}
where $B$ is the bag constant. The parameter $k$ depends on the mass of strange quark $m_s$ and the QCD coupling $\alpha_s$. For $m_s=0$, $k=1/3$, and for $m_s=250$ MeV/$c^2$, $k=0.28$. The bag constant $B$ takes a value $0.982B_0<B<1.525B_0$ for $m_s=0$, with $B_0=60$ MeV fm$^{-3}$  (\cite{N.Stergioulas}). In this paper, we take $B=60$ MeV fm$^{-3}$ (that is, $B^{1/4}\sim147$ MeV, corresponding to $\alpha_s=0$, as in Figure 1 of Ref.   \cite{MIT_jeff_B_value}). Moreover, in the ultra-relativistic approximation, we take $m_s=0$, so that $k=1/3$.

The above equations of state are displayed in Fig.  \ref{fig.eos}. The crossover between the MIT bag model and the unified equations of state are shown in the inset.

\section{Initial Conditions and Numerical Integration}\label{sec_ic}

To obtain the stellar structure of neutron stars and the metric potentials, we need to solve the TOV equations ( \ref{field equation}), ( \ref{TOV1}) and ( \ref{TOV2}) by numerical integration. We re-scale these equations by defining dimensionless quantities, $\eta=r/r_g$, $\tilde{P}=P/P_0$, and $\tilde{\rho}=\rho/\rho_0$, where $P_0=4B$, $\rho_0=P_0/c^2$ and $r_g=GM_{\odot}/c^2=1.4766\times10^{5}$ cm, the half of Schwarzchild radius of the Sun. 

We set the initial conditions in carrying out the numerical integrations as follows. Initially, we supply a value for the central density $\rho(0)=\rho_c$.  We note that $\lambda\rightarrow0$ and $\nu\rightarrow\nu_c$ as $r\rightarrow0$. Although it is straightforward to set the initial value $\lambda(0)=0$, it is not so straightforward to set the initial value $\nu(0)=\nu_c$ consistent with the initial choice $\rho(0)=\rho_c$. Since there is no direct dependence on $\nu$ in any of the equations, the central value $\nu_c$ is undetermined at this stage. However, it can be fixed so as to obtain an asymptotically flat solution at infinity. Consequently, the central value $\nu_c$ is fixed by looking at the exterior far-field solution, requiring it to be asymptotically flat so that $\lambda\rightarrow0$ and $\nu\rightarrow0$ as $r\rightarrow\infty$.

To validate this procedure, we solve equations ( \ref{field equation}), ( \ref{TOV1}) and ( \ref{TOV2}) employing the above initial conditions using different unified equations of state individually throughout the star, standard results for which are already available  (\cite{SLy_Chabanat_Haensel_1, bsk_Potekhin}). For each of these equations of state, we obtained the radial profiles of $\lambda(r)$ and $\nu(r)$ for the maximal central density corresponding to the maximal stable mass $M_{\rm max}$. We have confirmed that both $\lambda(r)$ and $\nu(r)$ asymptotically approach zero at infinity so that the space-time approaches the Minkowski flat geometry asymptotically.

The stellar mass of the star is obtained from
\begin{equation}\label{lambdamass}
 M=\frac{c^2}{2G}r_s\{1-e^{-\lambda(r_s)}\}
\end{equation}
where $r_s$ is the stellar radius, determined from the condition of pressure approaching zero.

Our calculation with the purely hadronic equations of state yield the maximal stable mass $M_{\rm max}$ to be 2.05, 1.81,01.86, 2.16 and 2.27 M$_\odot$ with SLy, FPS, BSk19. BSk20 and BSk21, respectively. Thses values agree well with previous calculations by \cite{SLy_Chabanat_Haensel_1} and \cite{ bsk_Potekhin} validating our numerical procedure. We also note that the maximal stellar mass $M_{\rm max}$ is obtained for central densities between $2\times10^{15}$ and $3.5\times10^{15}$ g cm$^{-3}$ across these hadronic equations of state.

However, just prior to attaining the maximal stable mass $M_{\rm max}$, BSk19 and BSk20 violate the condition of causality so that the velocity of sound given by $c_s=\sqrt{\frac{\partial P}{\partial \rho}}$ exceeds the velocity of light $c$. On the other hand, $M_{\rm max}$ given by SLy, FPS and BSk21 remain causal within the gravitational threshold of steability.

\section{Stellar Structure with Combined Equations of State}\label{sec_stellar_str}
As we have seen in the preceding section, some of the unified equations of state (BSk19 and BSk20) yield maximal masses by violating the condition of causality, occurring in the high density central region of the star. Moreover, the maximal mass was obtained for central densities between $2\times10^{15}$ and $3.5\times10^{15}$ g cm$^{-3}$. At such high densities, we expect that the core region to be composed of deconfined quark matter rather than hadronic matter. We thus take the MIT bag model to represent the deconfined quark matter in the core region. The region away from the core, on the other hand, is expected to be made up of predominantly hadronic matter that can be represented by one of the unified equations of state. This gives five combined equations of state, which we shall represent as MIT-SLy, MIT-FPS, MIT-BSk19, MIT-BSk20 and MIT-BSk21.

\begin{figure}[t!]
 \centering
 \includegraphics[width=0.8\textwidth, height=8cm]{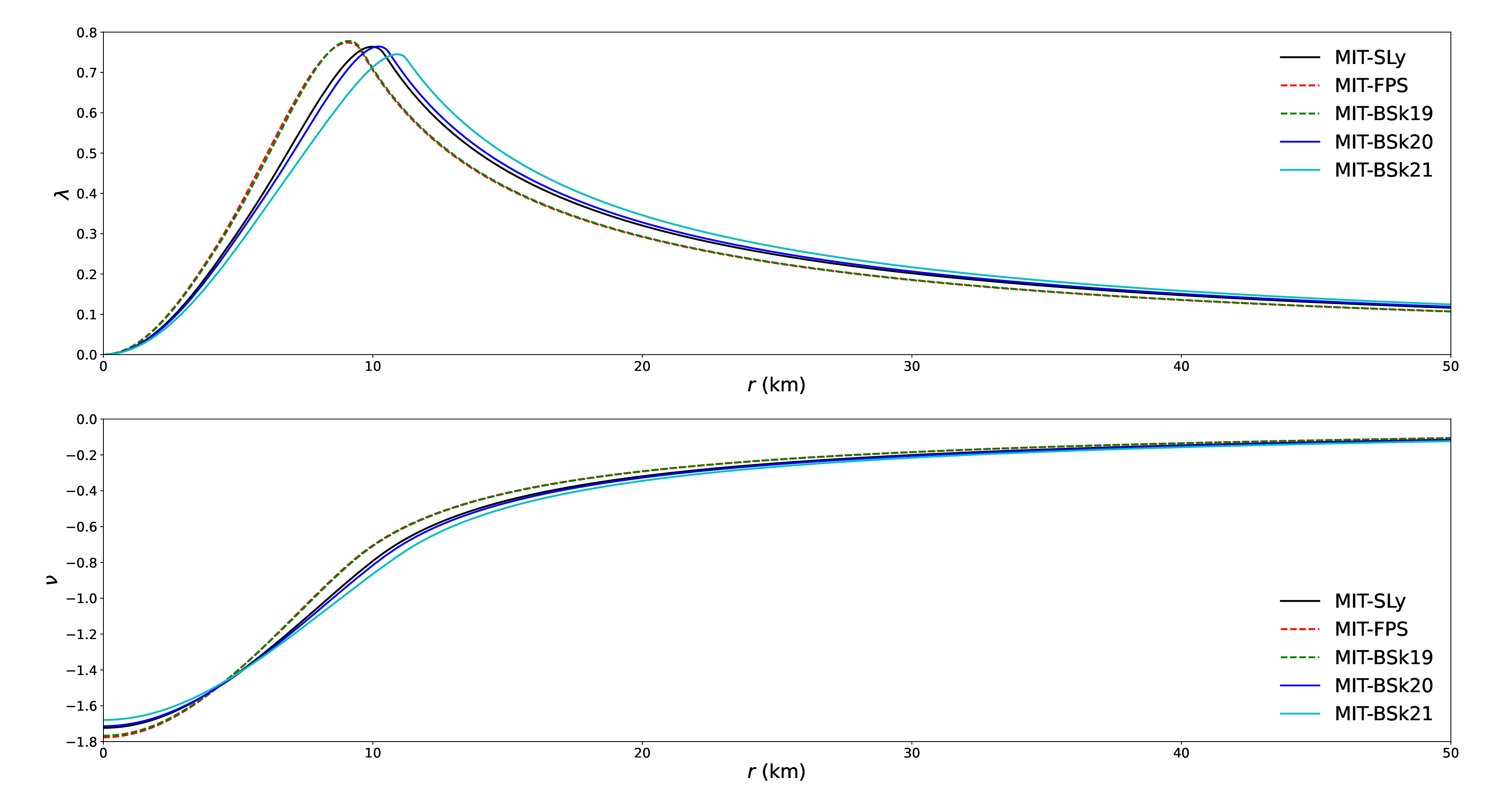}
 \caption{\footnotesize Radial profiles of the metric potentials $\lambda(r)$ [top] and $\nu(r)$ [bottom] for central densities corresponding to the maximal mass $M_{\rm max}$ using different combined equations of state.}
 \label{fig.lambdanu}
\end{figure}

\begin{center}
\begin{table}[b!]
\centering
\caption{\footnotesize Maximal stable mass $M_{\rm max}$, the corresponding central density $\rho_c$, and stellar radius $r_s$ for different equations of state. The table also shows the speed of sound $c_{\rm s,c}$ at the center and $r_{\rm peak}$ is the radial position of the peak in $\rho r^2$.\\}
{\small
\begin{tabular}{cccccc}
\hline
Equation  &$\rho_c$ ($10^{15}$& $r_s$&$M_{\rm max}$&$r_{\rm peak}$&$c_{s,c}$ ($10^{10}$ \\
of state  &g cm$^{-3}$)&(km)&($M_\odot$)&(km)&cm s$^{-1}$) \\
\hline
MIT-SLy   & $2.39$ &  10.75 & 1.85 & 7.89 & $1.73$ \\
MIT-FPS   & $2.94$ &   9.86 & 1.71 & 6.98 & $1.73$ \\
MIT-BSk19 & $2.87$ &   9.87 & 1.72 & 7.10 & $1.73$ \\
MIT-BSk20 & $2.28$ &  10.93 & 1.89 & 8.23 & $1.73$ \\
MIT-BSk21 & $2.03$ &  11.60 & 1.98 & 8.79 & $1.73$ \\
\hline
\end{tabular}}\label{tab.gr_equation of state}
\end{table}
 \end{center}


We take the crossover between the regions of deconfined quark matter and hadronic matter as the first intersection point between the two curves where both MIT bag model and the unified equation of state have the same pressure and the same density (shown in the inset of Fig.\  \ref{fig.eos}). This gives piece-wise continuous curves for each combination of equations of state. With these models, we numerically integrate the TOV equations given by ( \ref{field equation}), ( \ref{TOV1}) and ( \ref{TOV2}), employing the boundary conditions as described in Section  \ref{sec_ic}. This yields the interior and exterior solutions for the metric potentials $\lambda(r)$ and $\nu(r)$, and the radial profiles of density $\rho(r)$, pressure  $P(r)$ and mass $m(r)$ for different combinations of the equations of state.

We display in Fig.\  \ref{fig.lambdanu} the radial profiles of the metric potentials $\lambda(r)$ and $\nu(r)$ for different combinations of the MIT bag model with the unified equations of state corresponding to the maximal stable mass $M_{\rm max}$ (shown in Table\  \ref{tab.gr_equation of state}). These metric potentials display different behaviors up to $40$ km from the center and thereafter they vary in similar ways with respect to different equations of state. Moreover, we have confirmed that both $\lambda(r)$ and $\nu(r)$  approach zero at large values of $r$ signifying asymptotic approach to a flat Minkowski space-time at infinity.

\begin{figure}[t]
 \centering
 \includegraphics[width=.8\textwidth, height=7cm]{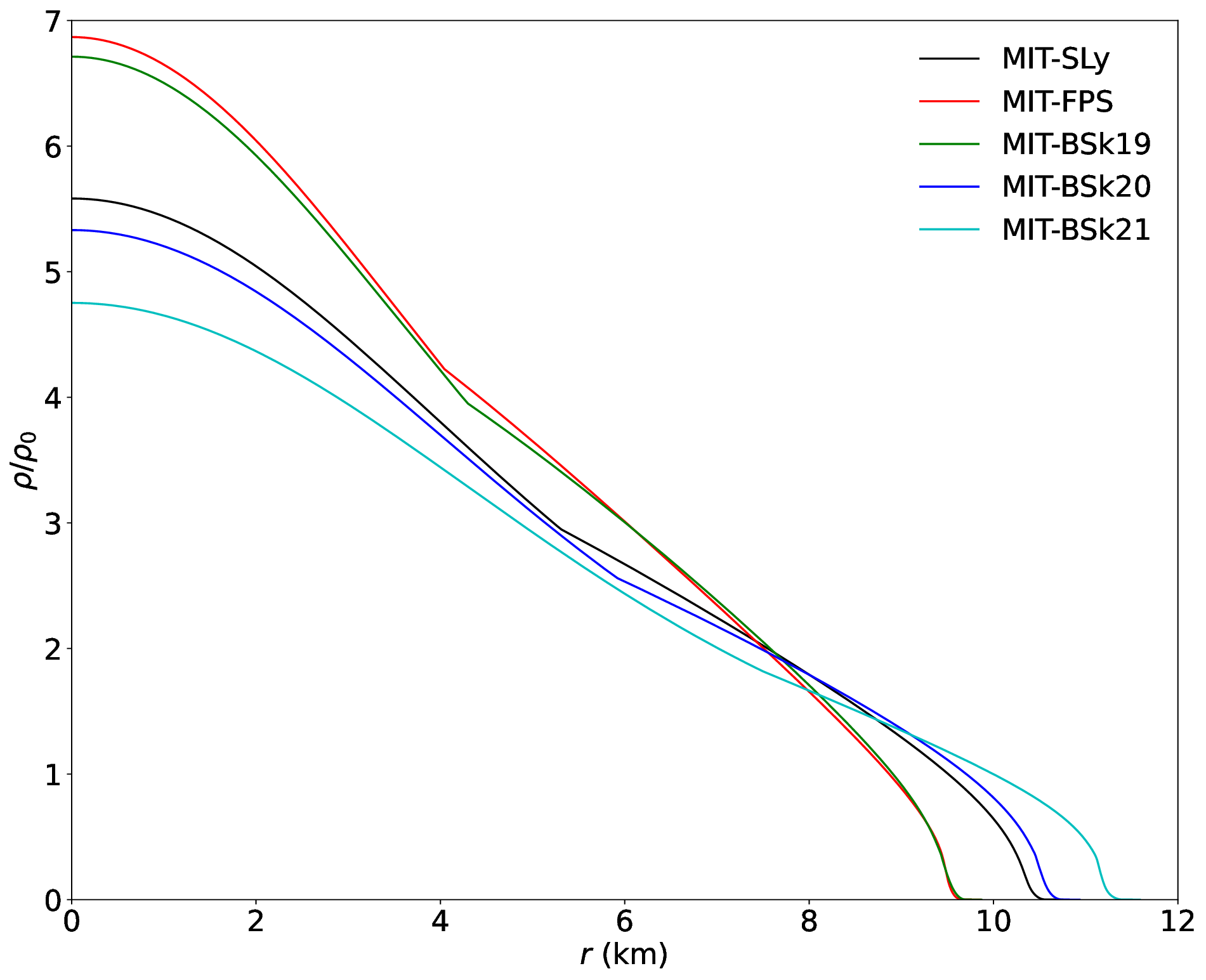}
 \caption{\footnotesize Density profiles $\rho(r)$ for central densities corresponding to the maximal mass $M_{\rm max}$ using different combined equations of state.}
 \label{fig.density}
\end{figure}

Fig.  \ref{fig.density} shows the density profiles corresponding to the maximum stable mass $M_{\rm max}$ for different combined equations of state. Although the MIT bag model represents the core in all cases, there are significant differences among the maximal central densities leading to different density profiles in all cases. Moreover,  the density profiles show significant variation in the region near the surface where different unified equations of state describe the hadronic matter.


Table\  \ref{tab.gr_equation of state} summarizes the results for the maximum stable mass $M_{\rm max}$ obtained with different combinations of equations of state. It is evident that, although the central region is governed by the MIT bag model, the maximum central density $\rho_{c,\rm max}$, the maximal mass $M_{\rm max}$ and the stellar radius $r_s$ differ for different combinations of equations of state. It is important to note that all these maximal central densities are consistent with the causality condition on the speed of sound, $c_s\leq c$.

\begin{figure}[t]
 \centering
 \includegraphics[width=.8\textwidth, height=7cm]{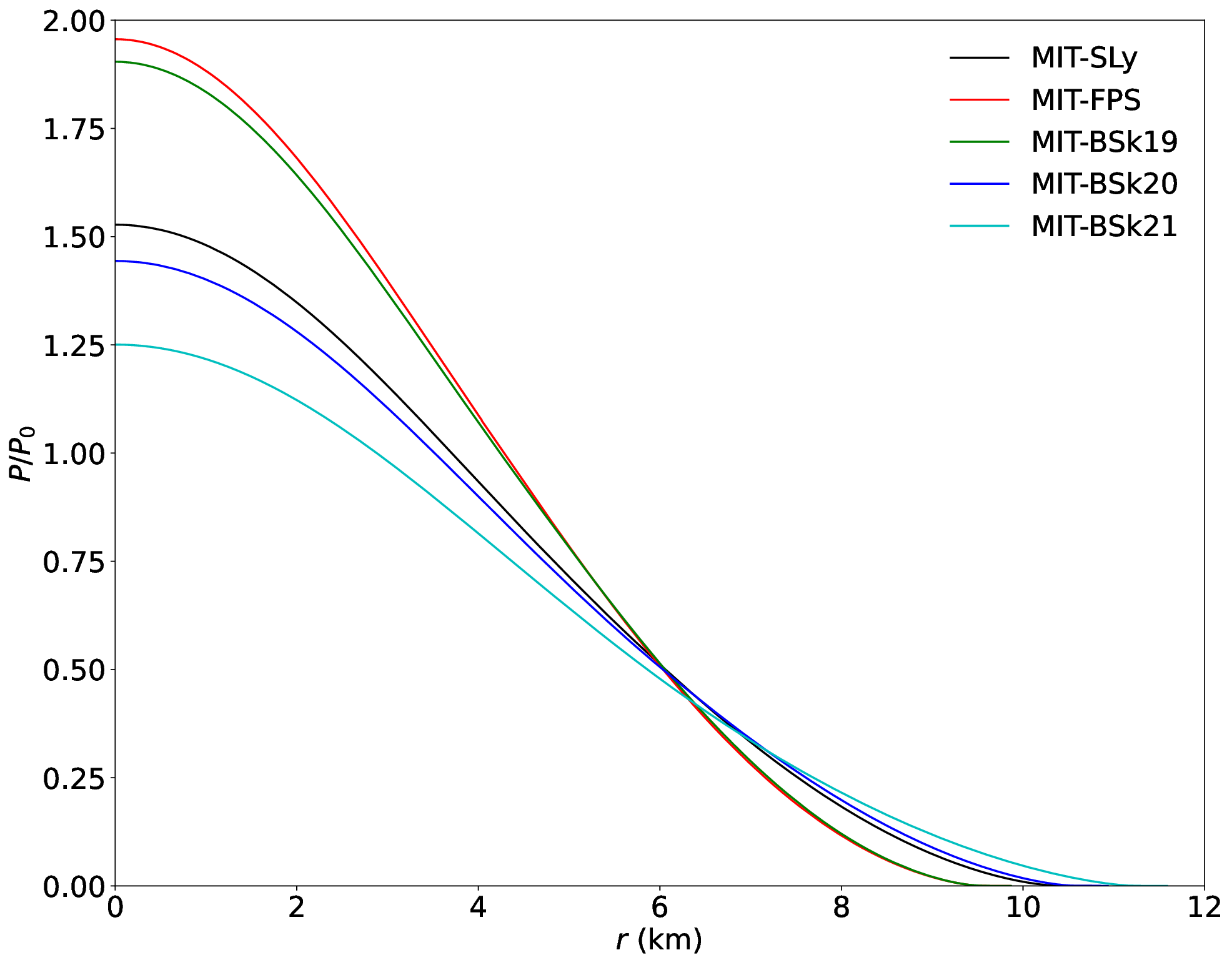}
 \caption{\footnotesize Pressure profiles $P(r)$ for central densities corresponding to the maximal mass $M_{\rm max}$ using different combined equations of state.}
 \label{fig.pressure}
\end{figure}

Fig.\  \ref{fig.pressure} displays the pressure profiles for the maximum stable masses with the five combinations of equations of state.  These profiles exhibit significant variation in the core region described by the MIT bag model, whereas there are small variations in the region near the surface described by different unified equations of state.

\begin{figure}[b!]
 \centering
 \includegraphics[width=.8\textwidth, height=7cm]{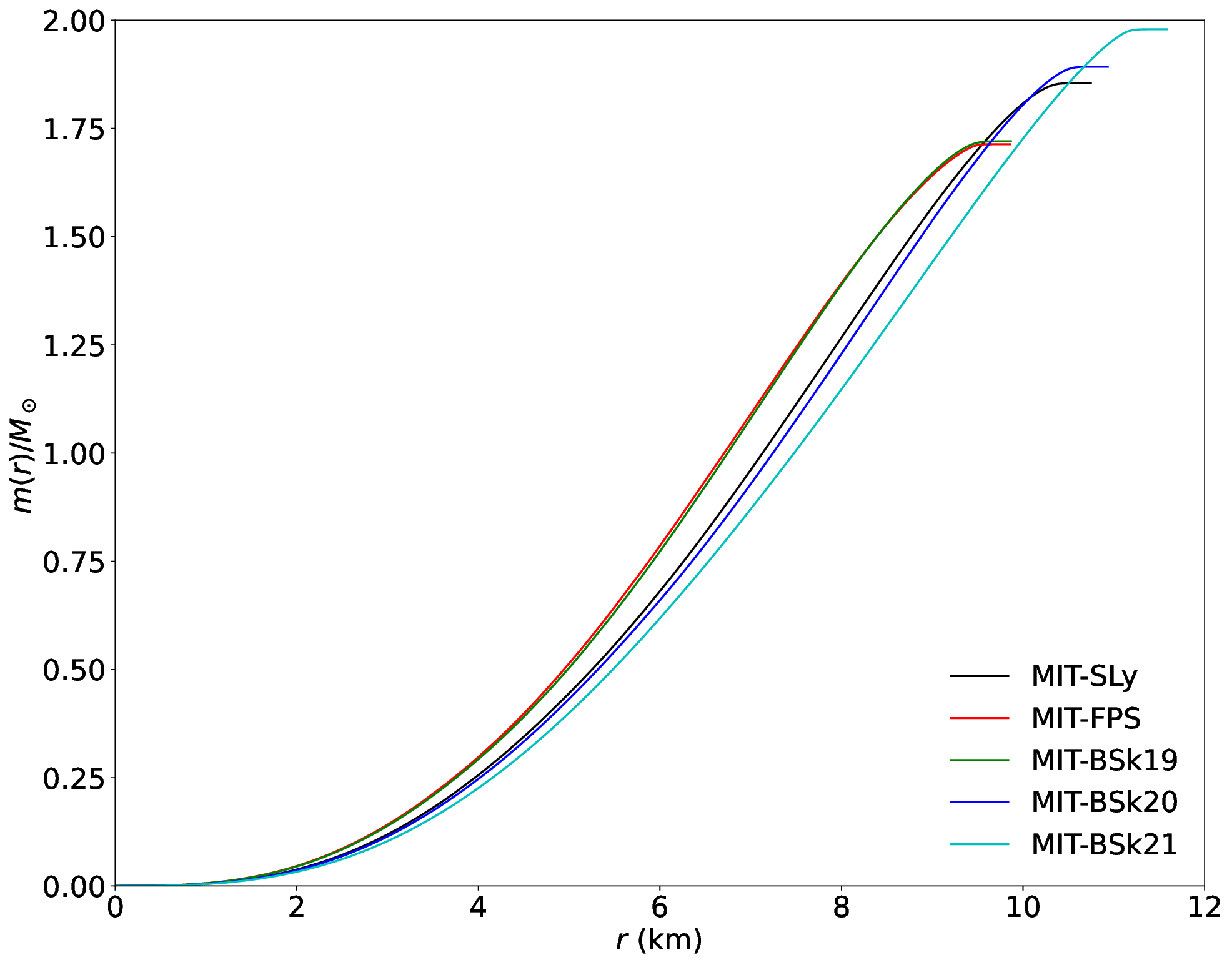}
 \caption{\footnotesize Radial profiles of the mass $m(r)$ corresponding to the maximal mass $M_{\rm max}$ with different combinations of equations of state.}
 \label{fig.massprof}
\end{figure}

 Fig.\  \ref{fig.massprof} displays the mass profile $m(r)$ given by equation ( \ref{mass_profile_eqn}) for the maximum stable masses $M_{\rm max}$ for different combinations of equations of state. The flattening of the curves near the surface occur due to negligible contribution to the total mass from near the surface.

 Unlike the cases with the individual equations of state, the maximum stable mass in the present combined cases are below $2$ M$_{\odot}$ as shown in Table\  \ref{tab.gr_equation of state}. For example, BSk21 alone gave a maximum stable mass of $2.27$ M$_\odot$ with a central density of $2.30\times10^{15}$ g cm$^{-3}$, whereas the combination MIT-BSk21 yields $1.98$ M$_\odot$ with a central density of $2.03\times10^{15}$ g cm$^{-3}$. Although the value $1.98\ M_\odot$ is obtained with a deconfined quark core, it is the closest to the value of $2$ M$_\odot$ among the combinations considered herein.

\begin{figure}[t!]
 \centering
 \includegraphics[width=.8\textwidth, height=7cm]{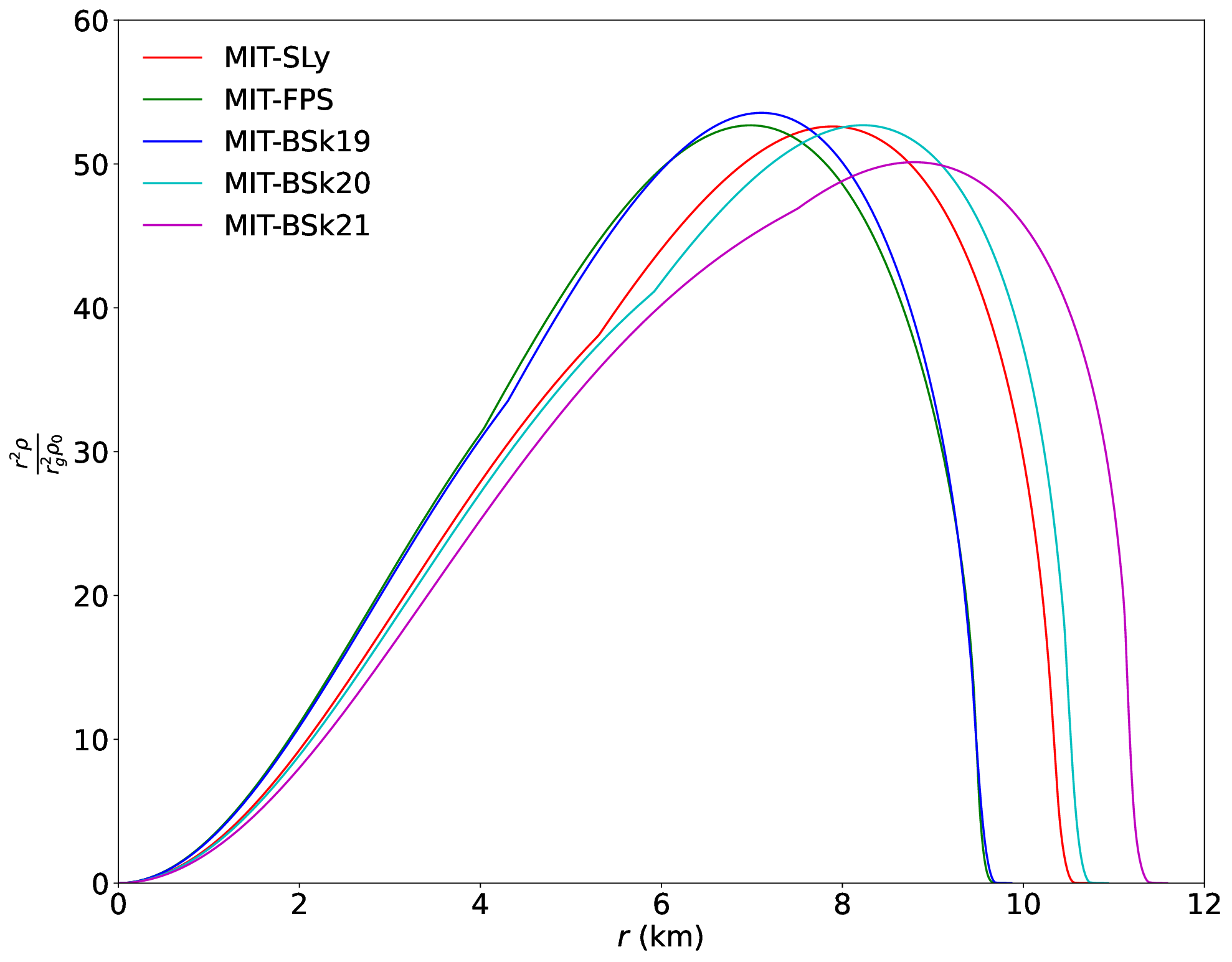}
 \caption{\footnotesize Radial profiles of $\rho r^2$ corresponding to the maximal mass $M_{\rm max}$ with different combinations of equations of state. These radial profiles illustrate the fact that the mass distribution moves to higher radial values as the peak value of $\rho r^2$ moves to the right, giving the highest total mass for the combination MIT-BSk21.}
 \label{fig.ver}
\end{figure}

 We further see from Table\  \ref{tab.gr_equation of state} that the maximal masses correlate well with the stellar radii. To understand this correlation, we plot in Fig.\  \ref{fig.ver} the profile of $\rho r^2$ for each of the combined equations of state. We note from Fig.  \ref{fig.ver} that the peaks in the radial profiles of $\rho r^2$, and hence the mass distribution, continues farther outwards for the combination MIT-BSk21 compared to the rest of the combinations, giving the highest maximal mass. The positions of these peaks are given in Table\  \ref{tab.gr_equation of state}. As the peaks shift to higher values of $r$, the mass distribution also shifts to higher values of $r$ so that the contribution to the total mass increases. This clearly indicates that the position of the peak is correlated with the maximal mass.

\begin{figure}[b!]
 \centering
 \includegraphics[width=.8\textwidth, height=7cm]{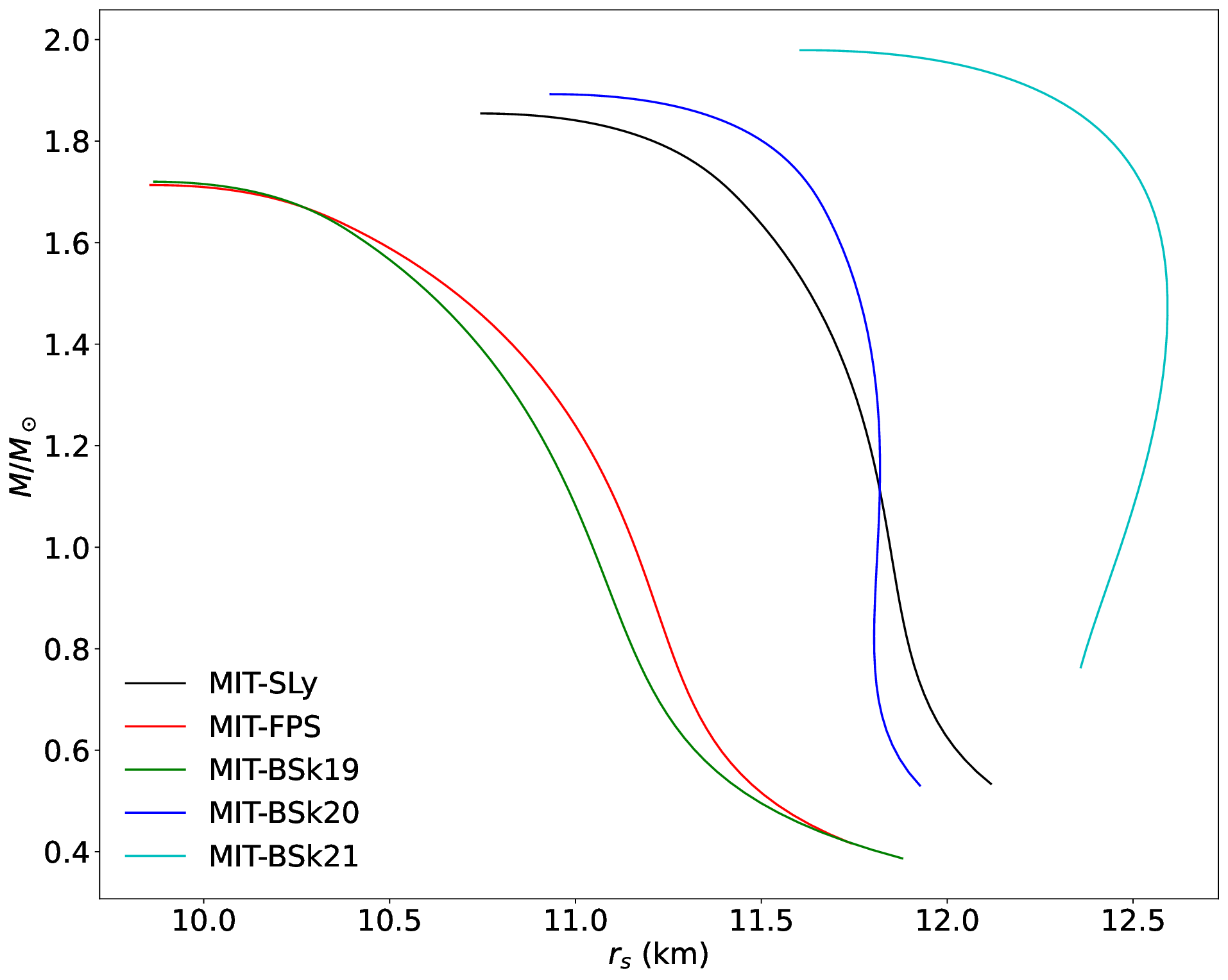}
 \caption{\footnotesize Mass-radius relations with different combinations of equations of state. Each curve is shown up to the corresponding M$_{\rm max}$ value.}
 \label{fig.MR_comb_equation of state}
 \end{figure}

 The contribution to mass $\Delta M_{12}$ contained between the radial coordinates $r_1$ and $r_2$, given by
 \begin{equation}\label{eqnmass2}
  \Delta M_{12}=4\pi \int_{r_1}^{r_2}\rho\,r^2 \,dr,
 \end{equation}
is determined by the area under the curve between $r_1$ and $r_2$ in Fig.\  \ref{fig.ver}. Thus the contribution to $\Delta M_{12}$ from the central region (between $0$ and $2$ km, say) is much smaller compared to the total mass in all cases. Moreover, contribution from near the surface is negligible as the curves fall almost vertically near the surface in all cases. Thus the dominant contribution to the total mass comes from the intermediate region surrounding the peak value of $\rho r^2$.

We further see from Fig.\  \ref{fig.ver} that the MIT-BSk21 peak (as well as its profile) extend up to a higher radial coordinate compared to the other four combinations. This is correlated well with the fact that MIT-BSk21 yields the highest maximal mass. In addition, we observe that MIT-FPS and MIT-BSk19 have nearly the same peak positions (and profiles), explaining nearly equal masses for them. Moreover, MIT-SLy and MIT-BSk20 have similar peak positions (and profiles), giving similar masses.

The above behaviors of increasing mass are correlated well with the shift in the position of the peaks towards higher radial coordinates (as shown in Table\, \ref{tab.gr_equation of state}). Since the integrand for $\Delta M_{12}$ given by equation ( \ref{eqnmass2}) is weighted by the factor $r^2$, it is the extent of the mass distribution that determines the maximal mass. Thus, it is the matter far from the center that gives dominant contribution to the total mass.

In the MIT-BSk21 combination, transition from the MIT bag model to BSk21 takes place at a lower density (higher radius) as compared to other combined cases. As a consequence, the MIT bag model dictates the stellar structure by forming a larger quark core region compared to the other combinations. In comparison, in the combinations MIT-BSk20 and MIT-SLy, the quark core is smaller, whereas in MIT-FPS and MIT-BSk19, the quark core is the smallest. In each combined case, a wide region around the peak is governed by the unified equation of state.

Fig.\  \ref{fig.MR_comb_equation of state} displays the mass-radius curves obtained for different combined equations of state. The maximum mass value in each of these curves corresponds to the maximal stable mass $M_{\rm max}$. The figure shows only the stable branches so that the curves stop at the $M_{\rm max}$ values. It is clearly seen that the combination MIT-BSk21 yields the highest maximal mass, and the MIT-FPS combination yields the lowest. The combinations MIT-FPS and MIT-BSk19 give nearly coincident mass-radius curves and maximal masses. The maximal masses are summarized in Table\  \ref{tab.gr_equation of state}.

\begin{figure}[t]
 \centering
  \includegraphics[width=.8\textwidth, height=7cm]{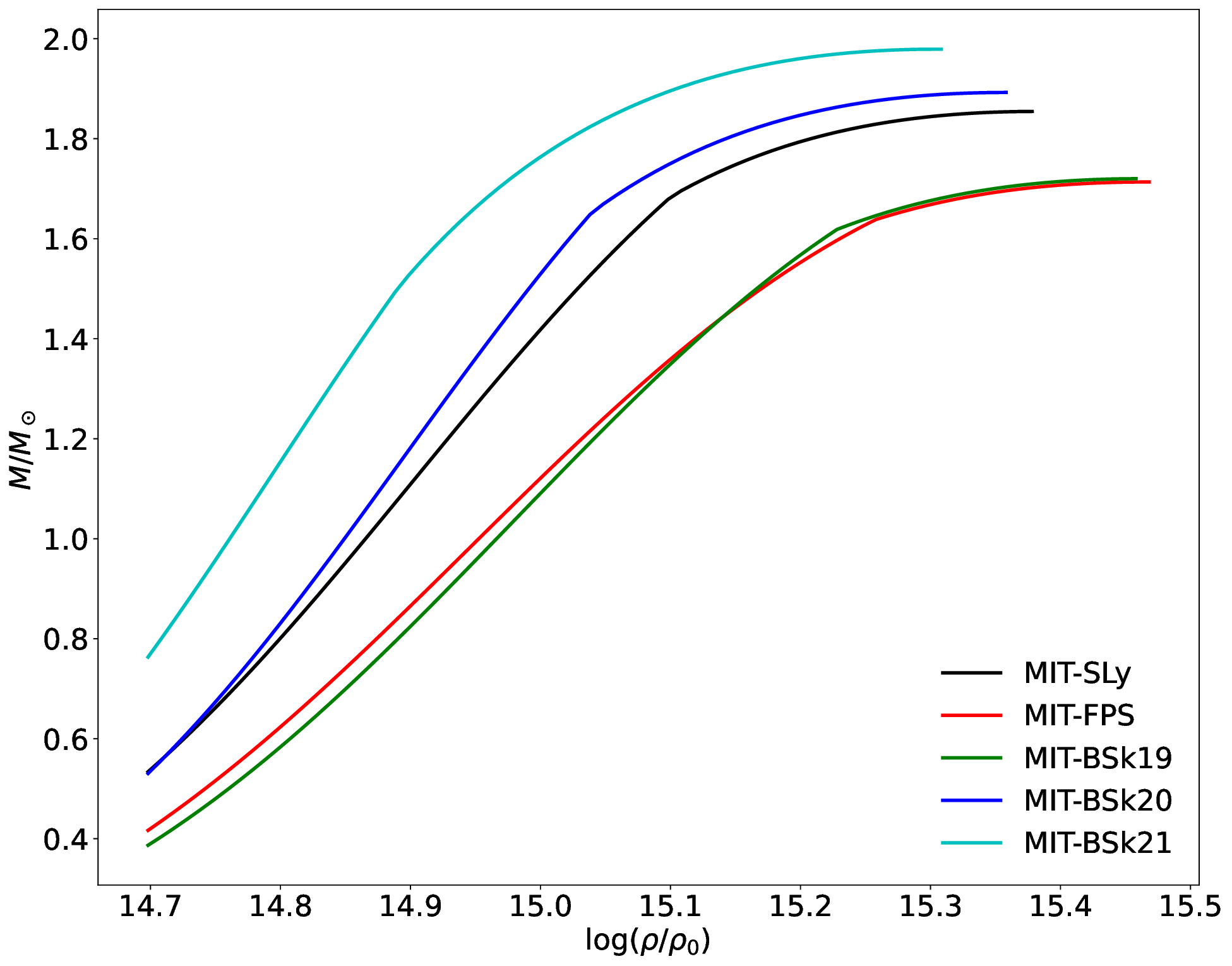}
 \caption{\footnotesize Mass versus central density for different combinations of equations of state. }
 \label{fig.MD_comb_equation of state}
\end{figure}

 Fig.\  \ref{fig.MD_comb_equation of state} displays the mass versus central density curves for different combinations of the MIT bag model with the unified equations of state. In each case, the maximum stable mass $M_{\rm max}$ corresponds to the maxima of the curve, where $\frac{\partial M}{\partial\rho_c}=0$. The region with the higher density segment beyond the maxima corresponds to the unstable regime, where $\frac{\partial M}{\partial\rho_c}<0$. The stable regime corresponds to the segment where the mass increases with increase in central density, that is, for  $\frac{\partial M}{\partial\rho_c}>0$.
 It is important to note that the maximum stable mass in each of these combined cases does not violate the causality condition ($c_s\leq c$), as shown in Table\  \ref{tab.gr_equation of state}

\section{Conclusions} \label{sec_concl}

In this work we studied the stellar structure and maximal mass of neutron stars using different combinations of equations of state. As we have discussed earlier, the relativistic equations of state, without involving quark deconfinement in the central region, have been employed throughout the star, yielding maximal mass $\sim 2\ M_\odot$ or even higher (Cf. Table III in Ref.  \cite{pollytrope_like_parametr}). However, as shown by \cite{pollytrope_like_parametr} some of the relativistic equations of state violate the condition of causality upon piecewise polytropic fits to the tabulated equations of state in the high density region.

When a unified equation of state is used throughout the star, we have seen in Section  \ref{sec_ic} that some of them yield the maximal mass $M_{\rm max}\sim 2\ M_\odot$ or higher. However, BSk19 and BSk20 violate the causality condition ($c_s\leq c$) in the high density central region for the maximal mass.

We also found that the central density corresponding to the maximal mass $M_{\rm max}$ is $\rho_c\sim 10\,\rho_{\rm nuc}$ or even higher, where $\rho_{\rm nuc}=2.67\times10^{14}$ g cm$^{-3}$ is the nuclear saturation density. This indicates that the matter is mainly deconfined quark matter in the central region. The MIT bag model is therefore a more suitable choice than the the unified equations of state in the central region. The outer region, on the other hand, may be assumed to be composed of mainly hadronic matter where one of the unified equations of state can be employed. The crossover between these two regions is determined by the matching point between the two equations of state. Since the crossover between the two phases happens in a small region, an accurate interpolation is not expected to make a significant difference in the {\em total} mass and a piece-wise continuous representation is expected to be reliable.

As shown in Table\  \ref{tab.gr_equation of state}, these combined equations of state yield maximal masses slightly less than $2\ M_\odot$, whereas, the MIT-BSk21 combination almost touches this value. In addition, we find that the causality condition ($c_s\leq c$) is well-respected for all these combined equations of state.

Our calculations illustrate another important feature determining the maximal mass. The integrand determining the total mass of the star is weighted by $r^2$ and the peak in $\rho r^2 $ occurs far from the center. Thus the dominant contribution to the total mass comes from the region around the peak. The regions near the center as well as the surface give small contributions to the total mass. The combination MIT-BSk21 yields the highest maximal mass as its peak in $\rho r^2$ occurs at the highest radial coordinate $r$ (among the combinations considered). It may also be important to note that the deconfined quark core for MIT-BSk21 extends to a higher radial coordinate than the other combinations. The peak in $\rho r^2 $ occurs in the outer hadonic matter region and it is moved to a higher radial coordinate for MIT-BSk21 than in the other combinations. The positions of the peaks in $\rho r^2 $ are well-correlated with the maximal mass as displayed in Table\  \ref{tab.gr_equation of state}.

 Recently, \cite{PSRJ095206072} estimated a pulsar mass of $2.35\pm0.17$ M$_\odot$ for  PSR J0952-0607 observed by \cite{PSRJ095206071} which is a rapidly rotating neutron star with pulsar frequency $707$ Hz. As shown earlier by \cite{20Percent_2} and \cite{20Percent_1}, a rapidly rotating neutron star can support  $20-25$ $\%$ more mass than the gravitational threshold of the non-rotating neutron star. Our gravitational threshold being 1.98 M$_\odot$ for a non-rotating neutron star, the increased threshold due to rotation would be $ 2.38-2.48$ M$_\odot$. This is consistent with the new benchmark of $2.35$ M$_\odot$.

Our study thus illustrates that it is possible to approach the desired value of $2\ M_\odot$ for a non-rotating neutron star with proper combination of equations of state for deconfined quark matter and hadronic matter so that the peak in $\rho r^2$ occurs at a sufficiently high value of the radial coordinate. The total mass being significantly dependent on the outer region where the peak in $\rho r^2$ occurs, an adequate formulation of the equation of state for hadronic matter is also important. In the realistic scenario, we expect a continuous phase transition from the deconfined quark matter in the core region to the hadronic matter in the outer region. This phase transition, which is not yet well-understood, needs an adequate formulation on the basis of quantum chromodynamics. This is however a formidable task and improvements in the neutron star models would require well-formulated phenomenological models accounting for the necessary features of such continuous phase transition.



\end{document}